\documentclass[aps,prd,twocolumn,groupedaddress,showpacs,floatfix]{revtex4}
\usepackage{amssymb}
\usepackage{amsmath}
\usepackage{graphicx}
\newcommand{\tabequitem}[1]{\raisebox{-.15cm}{#1}\raisebox{-.35cm}\mbox{}}
\begin{document}
\title{Late-time dynamics of brane gas cosmology}
\author{Antonio Campos}
\affiliation{Institut f\"ur Theoretische Physik,
             Universit\"at Heidelberg,
             Philosophenweg 16,
             D-69120 Heidelberg,
             Germany}
\date{\today}

\begin{abstract}
Brane gas cosmology is a scenario inspired by string
theory which proposes a simple resolution to the initial 
singularity problem and gives a dynamical explanation 
for the number of spatial dimensions of our universe.
In this work we have studied analytically and numerically
the late-time behaviour of these type of cosmologies 
taking a proper care of the annihilation of winding modes.
This has help us to clarify and extend several aspects of 
their dynamics.
We have found that the decay of winding states into 
non-winding states behaving like a gas of ordinary 
non-relativistic particles precludes the existence of a 
late expansion phase of the universe and obstructs the 
growth of three large spatial dimensions as we observe 
today.
We propose a generic solution to this problem by 
considering the dynamics of a gas of non-static branes.
We have also obtained a simple criterion on the initial 
conditions to ensure the small string coupling approximation 
along the whole dynamical evolution, and consequently, the 
consistency of an effective low-energy description.
Finally, we have reexamined the general conditions for a 
loitering period in the evolution of the universe which could 
serve as a mechanism to resolve the {\sl brane problem} - a 
problem equivalent to the {\sl domain wall problem} in 
standard cosmology -  and discussed the scaling properties of 
a self-interacting network of winding modes taking into account 
the effects of the dilaton dynamics.  
\end{abstract}
\pacs{04.50.+h, 98.80.Cq, 98.80.Bp}
\maketitle

\section{Introduction\label{sec:Intro}}
To understand the nature of the initial singularity and the 
origin of the number of dimensions of our universe are two
fundamental problems in cosmology.
Is it any physical law that allows the universe to avoid the 
initial singularity?  
Why we live in $3+1$ dimensions?
Is it, some how, possible to explain dynamically the spatial 
dimensionality of spacetime? 
In the standard theory of general relativity the number of
dimensions of the cosmological spacetime is not derived from 
any fundamental law but simply assumed to be four.
Moreover, it cannot address the problem of the initial
singularity either because it is quite reasonable to believe
that Einstein equations will not be valid close to the Planck 
scale.

An exciting potential resolution of these two issues within
the framework of superstring theory was proposed by 
Brandenberger and Vafa in the late 80's \cite{Brandenberger:1989aj}.
This proposal is based in a fundamental symmetry of string theory
called T-duality which states that the physics of strings in a 
box of size $R$ do not change if we replace the length of the box 
by $l^2_{st}/R$ (where $l^2_{st}$ is the string length).
This symmetry is not respected by the standard cosmological 
equations of general relativity but it is naturally implemented 
when the dynamics of a dilaton field is properly taking into 
account \cite{Tseytlin:1992xk,Tseytlin:1992ss}.

In this framework the background spacetime has the topology of a
torus with nine dimensions, which are assumed to be of equal size, 
and the dynamics is driven by a gas of fundamental strings.
The evolution of the universe is considered to be adiabatic, that 
is, there is no cosmological production of entropy, and, the 
string coupling constant is assumed to be small such that an 
effective tree-level approximation of the string theory is valid.
The string gas supports different string states which can be 
decomposed as combinations of oscillatory modes (stationary 
vibrating strings), momentum modes (non-stationary strings), and 
winding modes (strings wrapping around the torus).
The T-duality symmetry interchanges winding modes with momentum 
modes leaving the spectrum of the theory invariant under the 
inversion of the radius of the torus (the energy of oscillatory 
modes is independent of the size of the torus).
Under this symmetry the temperature of the string gas is also 
invariant in the sense that it is the same when the size of the 
universe is $R$ or $l^2_{st}/R$. 
That means that no physical singularity will occur as the radius 
of the torus is made indefinitely small ($R\ll l_{st}$) avoiding 
the inherent temperature singularity of standard cosmology.

This scenario also offers a mechanism for explaining an upper bound
for the spacetime dimensionality under the assumption of thermal 
equilibrium.
The net cosmological effect of winding modes is to stop expansion
even though their energy grows when the volume of space increases.
The reason is because they contribute to the expansion with a total 
negative pressure.
The evolution of winding modes can be though as equivalent to that
of large classical cosmic strings.
If thermal equilibrium is maintained energy from the winding states 
can be transferred to the rest of the states of the energy spectrum 
and we can keep the universe expanding.
The key observation is to realize that the probability of interaction
decreases with the dimensionality of spacetime.
As a result, the annihilation of strings with winding number can only 
be efficient if the number of dimensions of the spacetime is not 
larger than four.
Then, at some point in the evolution, six spatial dimensions of the 
torus must remain small while the other three become large relative 
to the string length.
Which dimensions grow and which stay small will mostly depend on 
thermal and quantum fluctuations. 
These qualitative arguments have been confirmed numerically 
\cite{Sakellariadou:1996vk,Cleaver:1995bw}.
Some thermodynamical aspects of a string gas and their implications 
for cosmology have been reviewed in \cite{Bassett:2003ck}.

Recently, these ideas have been revived by Brandenberger et 
al. \cite{Alexander:2000xv} in order to include extended 
degrees of freedom other than fundamental strings  
\cite{Polchinski:1995mt,Polchinski:1996na,Polchinski:1998}.
In this scenario the early universe is assumed to have $D+1$ 
spacetime dimensions with an isotropic toroidal topology 
and filled with a gas consisting of all possible branes with 
spatial dimension $p < D$ that the spectrum of a particular 
string theory could admit. 
Because Dp-branes respect T-duality \cite{Sen:1996cf} the 
initial singularity problem can also be easily resolved within 
this extended scenario \cite{Alexander:2000xv} (for a more 
detailed discussion see \cite{Boehm:2002bm}). 
As with fundamental strings, the energy of winding modes of 
Dp-branes grows with expansion, giving the larger contribution 
to the total energy of the gas those with the largest $p$, and
they also tend to prevent expansion.
Assuming thermal equilibrium as well, the cosmology of
brane gases may also provide an explanation for the dimensionality 
of spacetime by an analogous mechanism of self-annihilation of 
winding states.
Modes with larger $p$ are more massive and they must decay first
allowing only $2p+1$ spatial dimensions to grow.
The late-time evolution will be dominated by the winding modes
with the lower spatial dimension ($p=1$).
Then, a hierarchy of small dimensions is generated and the observed
dimensionality of our universe can be again explained.
Generalisations of brane gas models to manifolds with non-toroidal
topology and to anisotropic backgrounds have been studied in
\cite{Easson:2001fy,Easther:2002mi} and \cite{Watson:2002nx},
respectively.
Attends to extend and discuss these type of cosmological scenarios 
by including fundamental degrees of freedom of 11-dimensional M-theory 
have been given in \cite{Easther:2002qk,Alexander:2002gj}.

Even though the string considerations of \cite{Brandenberger:1989aj}
generalise quite easily to branes gases, they face a problem
similar to the standard {\sl domain wall problem} of cosmological
models admitting the spontaneous breaking of a discrete symmetry
in the early universe \cite{Zeldovich:1974uw,Vilenkin:1994}.
Causality indicates that despite an efficient annihilation of the 
winding states at least one Dp-brane across a Hubble volume should 
have survived. 
As with domain walls defects, the presence of even only one brane in 
our present horizon with an energy larger than the electroweak scale 
would have introduced fluctuations in the temperature of the microwave 
background radiation incompatible with current experimental bounds.
This observation pose a severe constraint on cosmological models filled
with branes gases.
A solution to this new {\sl brane problem}, proposed in 
\cite{Alexander:2000xv} and further developed in 
\cite{Brandenberger:2001kj}, is to invoke a sufficiently long period of 
cosmological loitering 
\cite{Lemaitre:1931,Glanfield:1966,Felten:1986,Sahni:1991ks,Feldman:1993ue}
which might have allowed the whole spatial extend of the universe to be 
in causal contact so the actual absence of branes can be explained by 
microphysical processes.

The purpose of this work is to clarify and extend previous
results on the late-time dynamics of brane gas cosmologies (BGC).
In particular, we have been interested to analyse several aspects 
of the cosmological evolution of this type of scenarios by 
incorporating the annihilation of brane modes with winding number 
appropriately.
First, we have discussed analytically the qualitative features 
which are not sensible to the details of the modelling of winding
mode decay.
We have found that spatial dimensions cannot grow large if the
brane states without winding number are produced in the form of 
ordinary non-relativistic matter.
We suggest that this obstruction to explain the dimensionality of 
the spacetime can be very easily resolved if a gas of non-static 
branes is considered.
Additionally, we have obtained a simple criterion on the initial 
conditions that guarantee the smallness of the string coupling 
at all times and, consequently, the consistency of an effective 
low-energy description for BGC. 
We have seen, by studying numerically how the dynamics change with 
respect to the values of some representative parameters, that the 
particular characteristics of the decay of winding modes mainly 
affect intermediate stages of the evolution of the universe. 
Finally, we have been interested to check the robustness 
of the resolution of the above mentioned {\sl brane problem} by
a phase of loitering and to investigate the scaling properties of 
a network of self-interacting winding states driven by the dynamics
of a dilaton field.

The rest of the paper is organised as follows. 
In Sec.~\ref{sec:BGC} we give a brief review of 
the main ingredients of BGC.
Sec.~\ref{sec:Late} is divided into two main parts.
In the first part we describe with some detail how the process of
winding mode decay into small loops without winding number
can be modelled.
The second part is devoted to investigate the corresponding 
late-time cosmological dynamics. 
The conclusions of our analysis are summarised in 
Sec.~\ref{sec:Conclusions}.

\section{Cosmology of brane gases\label{sec:BGC}}
The brane gas scenario assumes that the early universe is
filled with a gas in thermal equilibrium containing all the 
Dp-branes supported by the compactification of 11-dimensional 
M-theory on $S^1$.
All the nine spatial dimensions left after compactification
are considered to be toroidal and to start expanding adiabatically
with an initial size of the order of the string length.
The cosmological dynamics of this set up is dictated by  
the low-energy effective action, 
\begin{equation}
S_B
   = -  \int d^{D+1}x\, \sqrt{-G}\, \mathrm{e}^{-2\phi}
        \left[ R 
              +4 (\nabla_{\mu}\phi)^2
              -\frac{1}{12}H^2_{\mu\nu\alpha}
        \right], 
\end{equation}
where $D$ are the spatial dimensions, $R$ denotes the scalar 
curvature corresponding to the metric tensor $G_{\mu\nu}$, 
$\phi$ represents the dilaton field, which is related with the
radius of compactification, and $H_{\mu\nu\alpha}$ is the field 
strength of a bulk two-form potential $B_{\mu\nu}$. 
We are going to employ units in which the string length is 
$l_{st}\sim 1$.
The matter source in this scenario is given by a gas of 
Dp-branes. 
The dynamics of individual branes with $p$ spatial dimensions 
embbeded in a $(D+1)$-dimensional bulk spacetime is described 
by the Dirac-Born-Infeld action \cite{Polchinski:1998} 
(see also \cite{Leigh:1989jq}),
\begin{equation}
S_p
   = T_p \int d^{p+1}\xi\, \mathrm{e}^{-\phi}
       \sqrt{-\mathrm{det}( G_{\alpha\beta}
                           +B_{\alpha\beta}
                           +2\pi\alpha' F_{\alpha\beta})}, 
\end{equation}
where $T^{-1}_p=\sqrt{\alpha'}(2\pi\sqrt{\alpha'})^p$,
$\sqrt{\alpha'}$ is the string length $l_{st}$, the set of 
coordinates $\xi^\alpha (\alpha=0,\cdots,p)$ parametrise the 
D-brane world-volume, $G_{\alpha\beta}$ and $B_{\alpha\beta}$ 
are the pull-backs of the $(D+1)$-spacetime metric $G_{\mu\nu}$ 
and the antisymmetric tensor field $B_{\mu\nu}$, respectively, 
while $F_{\alpha\beta}$ is the field strength of a $U(1)$ gauge 
field $A_{\mu}$ living on the D-brane.
This action includes the dynamics of transverse mode fluctuations,
governed by $(D-p)$ world-volume scalar fields, and longitudinal 
mode fluctuations, described by the gauge field, in addition to 
winding modes giving the background mass of the brane.
In the standard picture of BGC it is assumed that brane mode 
fluctuations are small and only winding modes dominate the 
cosmological dynamics.
The mass energy of these modes is given by 
(see for instance \cite{Maggiore:1998cz,Boehm:2002bm}),
\begin{equation}
E^{(p)}_w = \tau_p\,  Vol_p.
\label{eq:brane_mass}
\end{equation}
with $\tau_p=T_p/g$ a re-scaled brane tension, 
$Vol_p$  the physical spatial volume of the brane, and
$g\equiv\exp(\phi)$ the string coupling which is considered to
be small. 
The equation of state corresponding to winding modes with
$p$ spatial dimensions is \cite{Alexander:2000xv,Vilenkin:1994},
\begin{equation}
P_w^{(p)}
   =\gamma_p E_w^{(p)}, 
   \textrm{\,\,\, with } \gamma_p=-\frac{p}{D}.
\label{eq:eq_state}
\end{equation}

In this work we shall restrict our analysis to spatially flat 
homogeneous and isotropic spacetime backgrounds,
\begin{equation}
ds^2
   = - dt^2
     + \mathrm{e}^{2\lambda(t)}\sum^{D}_{i=1}dx^2_i.
\end{equation}
In \cite{Watson:2002nx}, it was shown that the isotropy of the
three large spatial dimensions that remains after the decay of
the winding degrees of freedom comes out directly as a 
consequence of the dynamics.
Introducing the new dilaton variable $\varphi = 2\phi - D\lambda$, 
the background equations of motion derived from the total effective
action can be written as 
\cite{Tseytlin:1992xk,Tseytlin:1992ss,Veneziano:1991ek,Gasperini:2002bn},
\begin{eqnarray}
 \dot\varphi^2 - D\dot\lambda^2
&=& \mathrm{e}^\varphi E, 
\nonumber \\
 \ddot\varphi
-D\dot\lambda^2
&=& \frac{1}{2}\,\mathrm{e}^\varphi E,
\nonumber \\
\ddot\lambda
-\dot\varphi\dot\lambda
&=& \frac{1}{2}\,\mathrm{e}^\varphi P.
\label{eq:field_equations}
\end{eqnarray}
where $E$ and $P$ are the total energy and pressure
build on all (winding and non-winding) matter source 
contributions,
\begin{eqnarray}
E  &=& \sum_p E_w^{(p)}
      +E_{nw},
\\
P  &=& \sum_p \gamma_p E_w^{(p)}
      +\gamma E_{nw}.
\end{eqnarray}
For non-winding modes we will consider an ordinary 
equation of state $P_{nw}=\gamma E_{nw}$ with
$0\leq \gamma \leq 1$. 
The above dynamical equations obey an energy
conservation law, $\dot E + D\dot\lambda P =0$,
which comes out as a result of the adiabatic approximation
\cite{Tseytlin:1992xk}.

\section{Late-time dynamics\label{sec:Late}}
The early-time dynamics of BGC is quite well understood and it has 
been extensively discussed 
\cite{Tseytlin:1992xk,Alexander:2000xv,Bassett:2003ck,Tsujikawa:2003pn}.
In this work we are mainly interested in investigating the late-time
dynamics of this type of cosmologies.
For that purpose it is fundamental to supplement the field
equations given in (\ref{eq:field_equations}) with an appropriate 
description of the decay of winding modes into states without 
winding number.
We will assume that all winding states with $p>1$ have already been
completely annihilated and six spatial dimensions have been frozen.
Then, we can consider that the cosmological evolution is only driven
by the dynamics of winding modes with $p=1$ in three spatial
dimensions ($D=3$).

\subsection{Modelling the decay of winding modes\label{subsec:decay}}
The cosmological evolution of the winding strings can be thought 
as analogous to that of a network configuration of long cosmic 
strings in an expanding universe with toroidal topology 
\cite{Bennett:1986qt,Bennett:1986zn,Vilenkin:1994,Brandenberger:1994by}.
At time $t$ the total energy of a gas of $N_w$ winding
modes with individual mass-energy $\tau R(t)$, recall 
Eq.~(\ref{eq:brane_mass}), can be expressed as,
\begin{equation}
E_w(t) = N_w(t) \tau R(t) 
       = \tau l_c N_w(t)\mathrm{e}^{\lambda(t)}.
\label{eq:winding_energy}  
\end{equation}
The initial physical size of any of the spatial dimensions
of the torus, $l_c\exp(\lambda(0))$, is usually assumed to be 
of the order of the string length.
Using Eq.~(\ref{eq:winding_energy}) one can find an evolution 
equation for the total energy of the winding string gas,
\begin{equation}
\dot E_w 
   =  \dot\lambda E_w
     +\tau R(t) \dot  N_w(t).
\label{eq:winding_energy_evolution}
\end{equation}
The second term represents the loss of energy due to the
production of small loops without winding number.
If the winding mode decay is negligible the above equation
corresponds to the classical conservation equation of an 
ideal gas of non-interacting strings with equation of state 
$P_w=-E_w/3$, recall Eq.~(\ref{eq:eq_state}).
Long strings decay into small loop strings through 
self-interactions.
For such a network there exists a characteristic length scale 
$L_w(t)=R/\sqrt{N_w}$ which describes the typical separation 
between the string-like branes.
The rate of energy loss by the decay of winding modes into the
production of small loops can be roughly approximated by,
\begin{eqnarray}
\left. \dot E_w \right|_\mathrm{loop}
   &\sim& -c 
          \cdot
          \frac{1}{L^4_w}
          \cdot
          \tau'L_l
          \cdot
          V
\nonumber\\
    &\sim& -c\, \tau N^2_w(t) \left( \frac{L_l(t)}{R(t)} 
                         \right).   
\label{eq:decay_rate}
\end{eqnarray}
The parameter $c$ measures the efficiency of loop production,
the space volume is denoted by $V$, and $L_l(t)$ represents the 
typical size of the created loops.
The contribution $L^{-4}_w$ estimates the number of collisions 
per unit volume in a network with length scale $L_w$ whereas the 
second contribution, $\tau'L_l$, expresses the fact that the 
energy loss has to be proportional to the energy scale at which 
the small loops are produced.
In the last part of the above rate equation, we have assumed that
the string tension of the created loops should be similar to
that of the winding modes $\tau'\sim \tau$.
{}From Eqs.~(\ref{eq:winding_energy_evolution}) and 
(\ref{eq:decay_rate}) it is easy to obtain an evolution equation 
for $N_w(t)$,
\begin{equation}
\dot  N_w(t)
   = - \frac{c}{l^2_c} L_l(t) N^2_w(t) \mathrm{e}^{-2\lambda(t)}.
\label{eq:winding_energy_rate}
\end{equation}

Now it is convenient to introduce a dimensionless function of
cosmological time, $N_l(t)$, in order to write the total energy 
of small string loops as,  
\begin{equation}
E_l(t) = N_l(t) \tau R(t) \mathrm{e}^{-(1+3\gamma)\lambda(t)} 
       = \tau l_c N_l(t) \mathrm{e}^{-3\gamma\lambda(t)},
\label{eq:loop_energy}
\end{equation}
For $\gamma=0$ the produced loops behave like ordinary static
matter and for $\gamma=1/3$ as relativistic particles.
The function $N_l(t)$ describes the production of loops, it is
zero if there is no loops at all and constant if they are no 
longer created.
Thus, the rate of change for the total energy of produced loops 
will be,
\begin{equation}
\dot E_l 
   = -3\gamma\dot\lambda E_l
     +\tau R(t) \dot  N_l(t) \mathrm{e}^{-(1+3\gamma)\lambda(t)}.
\label{eq:loop_energy_evolution}
\end{equation}
The first term corresponds to the variation of energy due to
the expansion of the universe and the second to the energy
gained through winding mode decay.
Within our adiabatic approximation (recall that this assumption
is equivalent to the energy conservation equation discussed at the 
end of Sec.~\ref{sec:BGC}), we can find an evolution equation for 
$N_l(t)$,
\begin{equation}
\dot  N_l(t)
   =  \frac{c}{l^2_c} L_l(t) N^2_w(t) \mathrm{e}^{(3\gamma-1)\lambda(t)}.
\label{eq:loop_energy_rate}
\end{equation}
Obviously, $c=0$ corresponds to no loop creation.
Note that in this case the energy density of winding modes 
simply scales as $\rho_w\sim\exp (-2\lambda)$ and, consequently,
it should always dominate over ordinary matter or radiation 
components into the future of a expanding universe.

The field equations (\ref{eq:field_equations}) can now be 
cast into a close set of first-order differential equations
which completely describes the evolution of the universe 
during the process of winding mode decay ,
\begin{eqnarray}
\dot\varphi
&=& f, 
\label{eq:dot_varphi} \\
\dot\lambda
&=& l, 
\label{eq:dot_lambda} \\
\dot f
&=& 3l^2
   +\frac{1}{2}\,\mathrm{e}^\varphi 
    \left( E_w + E_l
    \right),
\label{eq:dot_f} \\
\dot l
&=& fl
   -\frac{1}{6}\,\mathrm{e}^\varphi 
    \left( E_w -3\gamma E_l
    \right),
\label{eq:dot_l} \\
f^2 
&=& 3l^2
    +\mathrm{e}^\varphi
     \left( E_w + E_l
     \right),
\label{eq:constraint}
\end{eqnarray}
together with Eqs.~(\ref{eq:winding_energy}), 
(\ref{eq:winding_energy_rate}), (\ref{eq:loop_energy}), and 
(\ref{eq:loop_energy_rate}).
Notice that $l$ is nothing but the Hubble parameter.

To complete the description of winding mode annihilation 
we will have to make an assumption about the formation of
small loops.
Following \cite{Bennett:1986zn} (see also \cite{Brandenberger:2001kj})
we will assume the loops are produced with a radius proportional
to cosmic time, $L_l(t) \sim t$.
This will imply that at late times loops are formed with a typical
size of the order of the Hubble horizon.
Another reasonable possibility would have been to assume that 
the radius of loops scales with the characteristic length of the
winding string network, that is $L_l(t) \sim L_w$.
We have, nevertheless, checked that both cosmological evolutions
are not substantially different.
All these points will be made more clear in a later section where 
we will discuss the scaling properties of the winding mode network.
Certainly, more sophisticated models to describe the decay
of winding modes could be conceived.
For instance, one can take into account the possibility of the
reconnection of small loops to winding strings and loop decay into 
gravitational radiation or ordinary matter (relativist or not).
However, for our purpose we can keep the description of winding
mode annihilation as simple as possible.

\subsection{Cosmological dynamics\label{subsec:dynamics}}
In the most simple cosmological picture of BGC the universe is assumed
to be initially in a state with all its spatial dimensions expanding 
isotropically, $l>0$, and with a physical size of the order of the 
string length, $l_{st}\sim 1$. 
As we have explained, these small toroidal dimensions can only 
become large if the winding and anti-winding modes can 
self-annihilate and decay into fundamental string loops or 
relativistic matter.
In what follows we will be interested to explore several dynamical
aspects of these cosmological scenarios.

\subsubsection{Qualitative analysis}
\begin{figure}[t]
\includegraphics*[totalheight=0.85\columnwidth,width=\columnwidth]{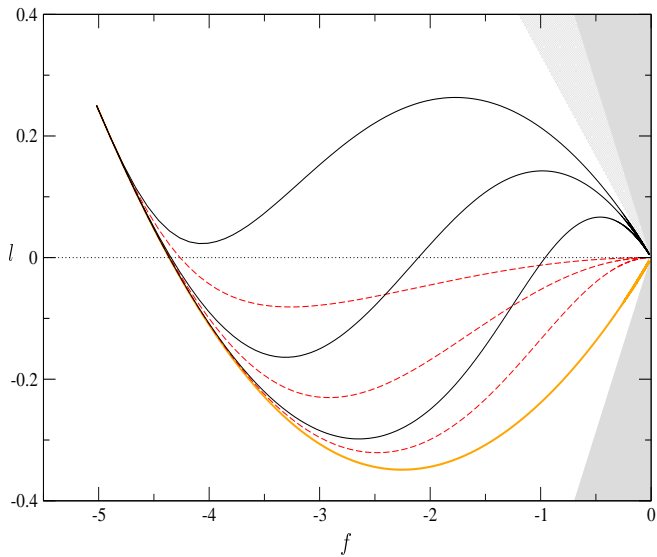}
\caption{Phase space for $(f,l)$.
In the dark grey area ($f^2-3l^2<0$) the total energy of the matter sources 
is negative and therefore it is excluded from the dynamical analysis.
The light grey dotted wedge, defined by the lines $l=-f/3$ and 
$l=-f/\sqrt{3}$, is a region where the smallness of the string coupling 
cannot be guaranteed.
We have plotted the numerical solutions of the equations of motion for 
several values of the physical parameters $c$, the efficiency of the
winding mode decay, and $\gamma$, the parameter characterising the
equation of state of the loops created.
The dashed lines correspond to solutions with $\gamma=0$ whereas the 
dark continuous lines to solutions with $\gamma=1/3$.
In both cases $c$ takes values $(0.1,1.0,10)$ from bottom to top.
For comparison we have also included the solution corresponding
to no winding mode decay $c=0$ (light continuous line).
\label{fig:phase_space}}
\end{figure}

Let us start this section illustrating some general properties
of the dynamical evolution of BGC including the effects of winding
mode decay but independently of the particular modelling of small 
loop creation.
Consider the two-dimensional phase space spanned by
$(f,l)$. 
The constraint equation (\ref{eq:constraint}) in conjunction with
the condition of positivity of the total energy restrict the
cosmological dynamics to values that satisfy the inequalities
\begin{equation}
f^2-3l^2 \geqslant 0
\,\,\,\,\, \mathrm{or} \,\,\,\,\,
|l| \leqslant \frac{1}{\sqrt{3}}|f|.   
\end{equation}
The straight lines $l=\pm f/\sqrt{3}$ correspond to solutions of
the equations of motion with zero total energy (that is $E_w=E_l=0$ 
except in the exceptional limiting case $\varphi\rightarrow -\infty$ 
where the total energy can take any finite positive value) passing 
through the origin. 
To probe the dynamical character of the fixed point $(f,l)=(0,0)$
it is sufficient to study the time evolution of these special 
solutions.
{}From Eq.~(\ref{eq:dot_l}) it is straightforward to check that
$l$ increases or decreases depending exclusively on the sign of $fl$.
In the first and third quadrant of the phase space $l$ must  
grow and in the second and fourth quadrant it has to diminish.
As a consequence, the two straight trajectories lying in the first 
and fourth quadrant move away from the origin and those in the 
second and third quadrant move closer to the origin.
Since self-consistency demands that there are no trajectories in
phase space crossing these special lines, any other trajectory
solution of the equations of motion originating in the half left 
side of phase space approaches the origin asymptotically whereas 
trajectories that start in the half right side diverge from it, 
see Fig.~\ref{fig:phase_space}.
Then, the fixed point $(f,l)=(0,0)$ is a saddle point and in fact
it is also, as it can be easily shown, the only critical point
of the equations of motion.
This simple dynamical picture is substantially modified if higher 
order curvature terms are included in the low-energy effective action 
\cite{Campos:2003}.

Now, let us see that for a given equation of state of the created 
loops all the solutions of the dynamical equations approach the 
origin asymptotically close to a particular straight line.
In other words, there exists a straight line which is a local solution 
of the equations of motion near the origin that attracts all other
dynamically allowed trajectories in phase space.
When the annihilation of winding modes is not taking into account
this straight line is $l=f/3$ \cite{Tseytlin:1992xk,Alexander:2000xv}, 
which is in fact a global solution of the equations of motion.
Intuitively one should expect that very close to the critical point
almost all winding modes have already disappeared and mainly the source
energy is in the form of a gas of string loops.
In this regime $f$ and $l$ obey two differential equations decoupled
from the rest of the variables,
\begin{eqnarray}
\dot f
&\simeq&  \frac{3}{2}l^2 
         +\frac{1}{2}f^2,
\nonumber\\
\dot l
&\simeq&   fl 
          +\frac{\gamma}{2}\left( f^2 -3l^2
                           \right).
\label{eq:close_origin}
\end{eqnarray}
To check our statement we have to look for straight line solutions 
inside the energetically allowed region, that is solutions of the 
form $l=\alpha f$ with $\alpha$ being a constant which obeys 
$|\alpha| \leq 1/\sqrt{3}$.
Substituting in the previous two equations we obtain a consistency 
algebraic equation for $\alpha$,
\begin{equation}
(\alpha + \gamma) (3\alpha^2 - 1) = 0.
\label{eq:special_lines}  
\end{equation}
Apart from the straight global solutions already studied 
($l=\pm f/\sqrt{3}$) we get a new solution which depends
on the equation of state of the produced loops  $l=-\gamma f$. 
For the static case we get the horizontal line $l=0$ and in the
relativistic limit the line $l=-f/3$. 
Obviously, this local solution does not exist if 
$|\gamma| > 1/\sqrt{3}$. 
To see that these new special lines are attractors of the
dynamics close to the origin we have to analyse the behaviour
of small deviations $l=\alpha f + \epsilon$ with $|\epsilon |\ll 1$.
Using Eqs.~(\ref{eq:close_origin}) again, the evolution of 
$\epsilon$ can be determined by the differential equation,
\begin{equation}
\dot\epsilon
   =  \left[ 1 - 3\alpha (\alpha + \gamma)
      \right]f\epsilon
     -\frac{3}{2}(\alpha + \gamma) \epsilon^2.
\end{equation}
For $\alpha = -\gamma$ this equation reduces to 
$\dot\epsilon = f\epsilon$ and then, noting that $f$ is negative
and cannot change sign, it is very easy to see that the absolute 
value of $\epsilon$ is always a decreasing function of time.
To probe the dynamical behaviour of the two other straight lines 
solutions looks much more subtle because the evolution of $\epsilon$ 
depends on the value of $\gamma$.
Moreover, second order effects will come to dominate at late times
and cannot be neglected.
In general, one can say that at early times the line 
$\alpha=+1/\sqrt{3}$ is an attractor of other trajectories and the 
line $\alpha=-1/\sqrt{3}$ is a repeller.
At late times all other solutions are attracted by both 
lines.

The relative behaviour of the rest of trajectories with respect 
to the special line $l=-\gamma f$ away from the origin is also
important to understand the global dynamics of the phase space.
In particular we are going to show that the solutions of the 
equations of motion can only cross the line from values of $l$
with $l>-\gamma f$ to values with $l<-\gamma f$ and never 
in the opposite direction.
As a consequence of this fact we will obtain two key dynamical 
properties of the equations of motion.
Let us start by considering the function defined by  
$\Phi(\varphi,\lambda;\alpha)\equiv\lambda - \alpha\varphi$,
which for $\alpha=-1/3$ is nothing but a rescaling of the original 
dilaton field $\Phi(\varphi,\lambda;-1/3)=(2/3)\phi$.
The special straight line is completely defined by extremizing this 
function $\dot\Phi(f,l;\alpha)=l-\alpha f=0$ for $\alpha=-\gamma$.  
Now consider a solution of the equations of motion crossing this
special line. 
If the trajectory cross from values of $(f,l)$ where $\dot\Phi > 0$
($l-\alpha f>0$) to values where $\dot\Phi < 0$ ($l-\alpha f<0$), 
then $\Phi(\varphi,\lambda;\alpha)$ will have a maximum.
To see that actually this is the only possibility and the crossing
cannot happen the other way round we have to compute  
$\ddot\Phi(f,\alpha f;\alpha)$ and check that it is strictly
negative.
Taking the evolution equations for $f$ and $l$ 
(\ref{eq:dot_f})-(\ref{eq:dot_l}) and using the constraint 
(\ref{eq:constraint}) we get, 
\begin{equation}
\ddot\Phi(f,\alpha f;\alpha)
   = \frac{\mathrm{e}^\varphi}{6}
     \left[  \left( 3\alpha-1
             \right) E_w
           +3\left( \alpha+\gamma
             \right) E_l
     \right],
\end{equation}
which for $\alpha=-\gamma$ simplifies to,
\begin{equation}
\ddot\Phi
   = - \frac{\mathrm{e}^\varphi}{6}  
       \left( 1+ 3\gamma
       \right) E_w.
\label{eq:crossing_condition}
\end{equation}
Thus, if the loops are produced with an equation of state
characterised by $\gamma > -1/3$ the negativity of 
$\ddot\Phi(f,\alpha f;\alpha)$ is always guaranteed.
In the special case $\gamma = -1/3$, which is completely equivalent
to consider a cosmological evolution without winding mode 
annihilation, the special line cannot be cross in neither direction.
This is nothing else but a reflection of the fact that $l=f/3$ is 
a global solution of the equations of motion.

It is worth to emphasise that the previous analytical result is 
independent of the evolution of the individual energies and thus of 
the equations of motion for the variables $N_w$ and $N_l$, and the 
particular modelling of the decay of winding modes. 
Moreover, there are two significant physical consequences that 
follow immediately.
First, consider the special line $l=-\gamma f$ with $\gamma = 1/3$.
Below this curve the dilaton and then the string coupling are always 
decreasing functions of time.
Then, if any trajectory in phase space, solution of the equations
of motion, starts with initial conditions below this curve one can 
ensure that the small string coupling approximation is dynamically 
preserved at all times as long as the equation of motion for the
produced loops obeys $\gamma \leqslant 1/3$.
And second, if the loops are produced in the form of ordinary 
non-relativistic matter, $\gamma=0$, and the universe enters a phase
of contraction it will never be able to re-expand.  
This simple analytical result is in contradiction with the solutions 
obtained in \cite{Brandenberger:2001kj}. 
Probably because they did not enforced the constraint equation 
(\ref{eq:constraint}) properly in their numerical analysis.
The late-time behaviour we have observed for $\gamma=0$ is very 
unsatisfactory from the phenomenological point of view because
it precludes the growth of three large spatial dimensions as we 
see today.
In a subsequent section we will discuss a possible way to resolve
this problem by considering a gas of non-static branes.

\subsubsection{Numerical analysis}
In the previous paragraph we have discussed some properties of
the dynamics of BGC which are independent of how the decay
of winding modes is described by the equations of motion and we 
have shown that the final fate of the such cosmologies only 
depends on the equation of state of the produced loops.
Now we are interested in studying numerically how this modelling 
influence and determine the intermediate evolution.
For that purpose we have use a fourth-order Runge-Kutta method
\cite{Gerald:1984} to solve Eqs.~(\ref{eq:winding_energy_rate})
and~(\ref{eq:loop_energy_rate})-(\ref{eq:dot_l}) for several
representative values of the parameter $c$, the efficiency of
winding mode annihilation, and $\gamma$, the parameter characterising 
the equation of state of the small loops created.
When solving numerically the equations of motion for this problem one
has to be very careful and use an efficient numerical algorithm with 
a sufficiently small spacing for the mesh because as the light grey 
dotted wedge depicted in Fig.~\ref{fig:phase_space} is approached the 
solutions are highly sensitive to numerical errors.
This explains why the authors of \cite{Brandenberger:2001kj} could 
have incorrectly found a runaway solution crossing the vertical axis 
($f=0$) which is completely inconsistent with the original equations
of motion.

Our previous qualitative analysis permits to constraint possible 
dynamically interesting, from the cosmological point of view, 
initial conditions.
Our small universe has to be initially expanding, thus we have to take
a positive $l(0)$.
Since we are not interested to be very close to the special lines
with zero total energy to enforce the small string coupling approximation
one cannot take a very negative initial $\varphi$.
The initial values we have chosen for all our numerical simulations
are $\varphi(0)=0$, $l(0)=0.25$, $N_w(0)=25$, $N_l(0)=0$.
The constraint (\ref{eq:constraint}) has been used to obtain the initial
value for $f(0)$.
The starting strength of the string coupling we have taking 
is $g(0)=0.1$ which together with $\varphi(0)$ permits to determine 
$\lambda(0)=(2\ln g(0)-\varphi(0))/3$.
Finally, we have taken $\tau=1$ and fixed the parameter $l_c$ using the
condition $l_c\exp{(\lambda(0))}\sim 1$.
It is also important to recall at this point that all dimensional 
quantities are measured in units of $l_{st}$.
It is easy to check that the small coupling approximation will 
always be guaranteed with this choice of initial conditions.

\begin{figure}[t]
\includegraphics*[totalheight=0.85\columnwidth,width=\columnwidth]{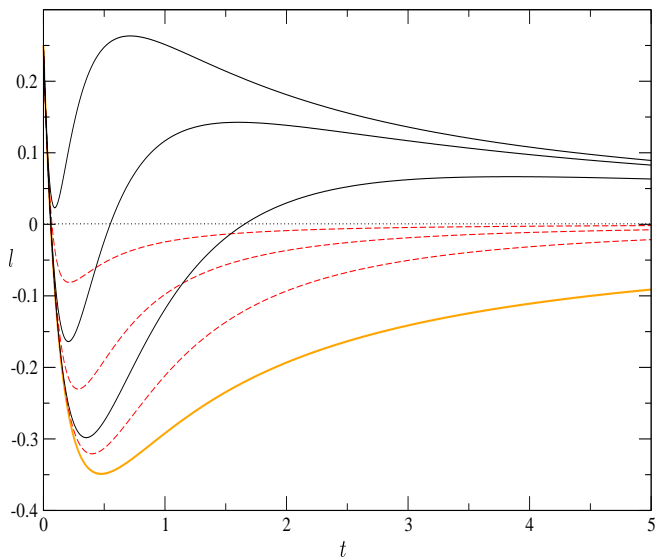}
\caption{Hubble parameter $l=\dot\lambda$ as a function of cosmic 
time.
The plotted curves represent solutions of the equations of motion 
with parameters $c$ and $\gamma$ chosen in the same manner as in 
Fig.~\ref{fig:phase_space}.
\label{fig:hubble}}
\end{figure}

\begin{figure}[t]
\includegraphics*[totalheight=0.85\columnwidth,width=\columnwidth]{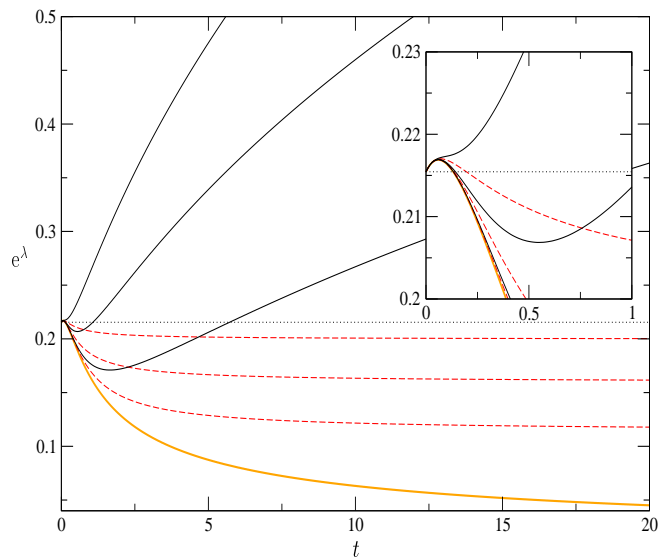}
\caption{Scale factor as a function of cosmic time.
The plotted curves represent solutions of the equations of motion 
with parameters $c$ and $\gamma$ chosen in the same manner as in 
Fig.~\ref{fig:phase_space}.
\label{fig:scale}}
\end{figure}

In Fig.~\ref{fig:phase_space} we have plot the phase space for the 
variables $(f,l)$, that is $(\dot\varphi,\dot\lambda)$.
The dark grey area is the region forbidden by the condition of 
positivity of the total energy of the system ($f^2-3l^2>0$)
and therefore it is excluded from the dynamical analysis.
The white area and the light grey dotted wedge are regions allowed
by energetic considerations.
The light grey dotted wedge is defined by the lines $l=-f/3$ and 
$l=-f/\sqrt{3}$ and is the region where the string coupling grows and 
its smallness cannot be guaranteed.
As emphasised before, trajectories originating in the white area can never 
cross into this region if $\gamma \leqslant 1/3$.
The dashed lines correspond to solutions with $\gamma=0$ (static loops) 
whereas the dark continuous lines to solutions with $\gamma=1/3$
(relativistic loops).
In both cases $c$ takes values $(0.1,1.0,10)$ from bottom to top.
For comparison we have also included the solution corresponding
to the case in which the winding modes do not self-annihilate $c=0$ 
(light continuous line).
The evolution of the Hubble parameter and the scale factor as a function 
of cosmic time has also been plotted in Fig.~\ref{fig:hubble} and 
Fig.~\ref{fig:scale}, respectively.

As one can immediately see from the numerical solutions studied
the efficiency parameter $c$ only qualitatively influences the very 
early dynamics of the equations of motion.
In general, as it grows the maximum contraction rate decrease for
any value of $\gamma$ (see Fig.~\ref{fig:hubble}) and if it is
sufficiently large the universe could never enter a phase of 
contraction. 

As expected from our previous qualitative analysis the very late-time
cosmological dynamics of this model is mostly independent of the 
parameters which characterises the decay of winding modes. 
In fact, the scale factor very soon behaves like,
\begin{equation}
\mathrm{e}^\lambda
   \sim t^{2\gamma/(1+3\gamma^2)},  
\end{equation}
which corresponds to a radiation-dominated universe for $\gamma=1/3$
and to a flat static universe for $\gamma=0$.
It is interesting to note that the above expansion law only coincides 
with that of the standard cosmological scenario for $\gamma=1/3$.
This is because as the winding modes self-annihilate all trajectories 
approach asymptotically the line $l=-f/3$ where $\dot\phi\sim 0$ and 
$\ddot\phi\sim 0$.
Thus, the original dilaton freezes dynamically without the need of 
introducing a dilaton potential and a standard expansion law is 
recovered before reaching the point $(f,l)=(0,0)$.
This is not the case if $\gamma\neq 1/3$.

With $\gamma=0$ and for any reasonable value of the efficiency 
parameter $c$ we find that the universe always enters a phase of 
contraction after a very short period of time and never re-expands 
despite the self-annihilation of winding modes 
(see Figs.~\ref{fig:phase_space}~and~\ref{fig:hubble}).
Thus if the winding modes decay into static loops the spatial
dimensions can never become large and the dimensionality of our
spacetime cannot be explained.
Apart from our previous qualitative arguments this result can also be 
physically understood in the following way.
Eq.~(\ref{eq:dot_l}), viewed as an equation for $\lambda$, is equivalent
to an equation for a damped point particle moving in an an effective 
potential with the following form,
\begin{equation}
U_{eff}(\lambda)
   = \frac{\tau l_c}{6}
     \mathrm{e}^\varphi
     \left[ N_w(t)\, \mathrm{e}^\lambda
           +N_l(t)\, \mathrm{e}^{-3\gamma\lambda}
     \right].
\end{equation}
In the particular case in which $\gamma=0$, the second term is no longer 
a function of $\lambda$ and it can be merely interpreted as a time 
modulation of the potential origin.
For such a potential, a universe that starts expanding will reach a 
maximum size and, at some point, it might enter a contraction phase from 
which it will not be able to escape.
As we have seen numerically the effect of the modulating factor 
$\exp(\varphi)$, which is always a decreasing function of time, is not 
relevant for this argument to hold.
This picture changes drastically at late times if $\gamma >0$
(See Figs.~\ref{fig:hubble}~and~\ref{fig:scale} for the case
$\gamma=1/3$).
Now the second term in the effective potential $U_{eff}$ comes 
to dominate the dynamics as the number of winding modes $N_w$ 
goes to zero and $N_l$ grows.
Thus, if a universe that started in a expanding phase passes through
a stage of contraction it will inevitably go back to a new period 
of expansion.
As a conclusion, we have given analytical and numerical evidences to
the fact that, in contrast to the numerical results of 
\cite{Brandenberger:2001kj}, 
re-expansion in the cosmological evolution of the universe is only 
possible if the small loops produced do not behave like ordinary static 
matter.

\begin{figure}[t]
\includegraphics*[totalheight=0.85\columnwidth,width=\columnwidth]{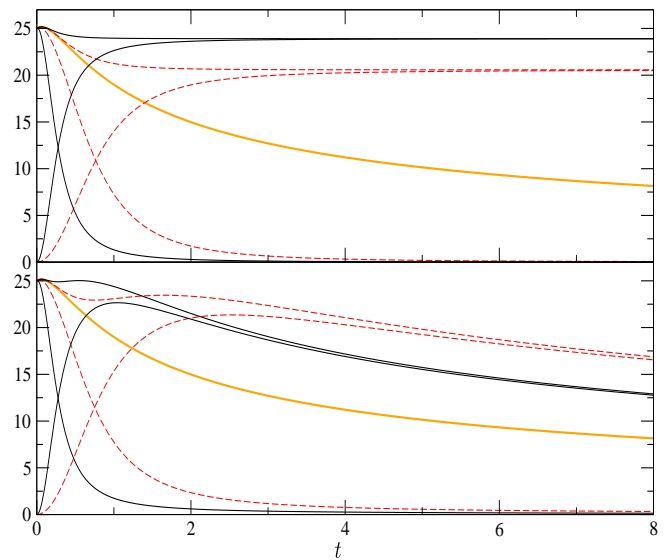}
\caption{Energy profiles for $\gamma = 0$ (top) and $\gamma = 1/3$ 
(bottom). 
In both plots the dark lines correspond to $c=1.0$, the light dashed 
lines to $c=0.1$, and the light continuous line to $c=0$, that is, the
case without winding mode decay (Recall that in this particular case 
$E_l=0$ and then the total energy is equal to $E_w$). 
The curves that asymptotically approach the zero axis represent the
energy of winding modes and the curves starting at zero the energy
of the loops produced.
The curves tracked at late times by the loop energy is the total
energy of the system $E_l+E_w$.
\label{fig:energy}}
\end{figure}

In Fig.~\ref{fig:energy} we show the profiles of all the energy 
components for $\gamma = 0$ and $\gamma = 1/3$. 
In both cases, the energy of winding modes goes to zero very 
rapidly and, consequently, the energy of small loops tracks 
the total energy of the universe during most of the time 
evolution.
One can see, in general, that the small loop energy very soon
evolves as,
\begin{equation}
E_l
   = \tilde{E}_l\mathrm{e}^{-3\gamma\lambda},  
\end{equation}
where $\tilde{E}_l$ is a constant which depends on the parameters
of the decay process.
For $\gamma = 0$ that means that the total energy reaches a
constant value.
We have checked numerically that the asymptotic value of this 
constant grows with an increasing efficiency parameter $c$.
In ordinary Einstein theory a constant energy would have meant
a matter-like dominated expansion of the universe.
However, in our brane gas model, the effect of the dilaton 
coupling drives the universe through an ever contracting phase.
On the other hand, for $\gamma=1/3$ we have a radiation-dominated 
like cosmological expansion for the universe.   
In this particular case the total energy decreases with time
more rapidly the bigger is the parameter $c$.
This is mainly due to the fact that the expansion goes faster
if $c$ is larger.

\subsubsection{Loitering and the {\sl brane problem}} 
The hierarchical picture of Dp-brane decay explains the actual 
dimensionality of spacetime though it leads to a variant of the 
domain wall problem in cosmology. 
A solution to this problem is to invoke a stage of
cosmic inflation before the branes can dominate the energy 
density of the universe. 
An alternative proposed in \cite{Alexander:2000xv}, and further 
developed in \cite{Brandenberger:2001kj}, was to advocate a 
late phase of loitering in the universe. This is a stage in
which the universe halts and its spatial extend $R$ could
become much smaller than the Hubble horizon $d_H\sim l^{-1}$,
that is a period in which $Rl\ll 1$.

We have checked numerically that a sufficiently long period in which 
this condition holds is quite natural in BGC, 
see Figs.~\ref{fig:hubble} and \ref{fig:scale}.
At early times and as long as the efficiency parameter $c$ is not too 
large one can assure that the Hubble horizon always becomes extremely 
large allowing the whole universe to be in causal contact.
However, it is not hard to show that at late times one has, 
\begin{equation}
Rl
   \rightarrow 
         \left\{ 
               \begin{array}{ll}
                 0 & \textrm{\,\,\, for } \gamma=0, \\
                 t^{-1/2} & \textrm{\,\,\, for } \gamma=\frac{1}{3},
               \end{array} 
         \right.
\end{equation}
and then, even if the early period of loitering is not long enough 
the Hubble horizon always becomes larger than the 
spatial extend of the universe and the {\sl brane problem}
can be solved.

\subsubsection{Scaling}
An important characteristic of the evolution of a self-interacting
network of strings in a expanding universe is that it reaches a stage
in which its characteristic length remains constant with respect to 
the size of the Hubble horizon $d_H$
\cite{Bennett:1986zn,Bennett:1988vf,Bennett:1989ak,Bennett:1990yp}.
An interesting question that we can answer with our analysis is whether 
this scaling regime is also reached when the cosmological dynamics is 
driven by the effects of a dilaton field.

When the evolution of the string network is assumed not to affect the
standard (say, radiation- or matter-dominated) expansion of the universe,
it makes no difference to talk about scaling with respect to the
Hubble horizon or to the cosmic time because they are always
proportional $d_H\sim l^{-1}\sim t$.   
However, when the back reaction on the dynamics of the gravitational 
background is taking into account the distinction between these two 
quantities become important because, as it happens in our problem, 
$d_H$ could become extremely large at some finite time ($l\rightarrow 0$). 

To study the scaling properties of string configurations it is usually 
convenient to introduce a dimensionless parameter $\zeta(t)=t/L_w(t)$ 
measuring how far is the system from a scaling behaviour relative to
cosmic time. 
The scaling regime is reached if $\zeta(t)$ relaxes to a constant 
value $\zeta_\ast$.
In this situation, the decay of the number of winding modes can be 
expressed as,    
\begin{equation}
N_w
   =    \zeta^2_\ast
        \left( \frac{l_c}{t}
        \right)^2 \mathrm{e}^{2\lambda}
   \sim \left\{ 
               \begin{array}{ll}
                 t^{-2} & \textrm{\,\,\, for } \gamma=0, \\
                 t^{-1} & \textrm{\,\,\, for } \gamma=\frac{1}{3}.
               \end{array} 
         \right.
\label{eq:scaling_decay}
\end{equation}
which yields an energy density of winding modes that decrease with 
time independently of the equation of state of the loops created,
\begin{equation}
\rho_w
   = \zeta^2_\ast 
       \left( \frac{\tau}{t^2}
       \right).  
\end{equation}
By substituting Eq.~(\ref{eq:scaling_decay}) into the dynamical equation 
for $N_w$ given in~(\ref{eq:winding_energy_rate}), it can be seen that
the parameters of a scaling solution must obey the following condition,
\begin{equation}
c
   = \frac{2(1-2\gamma+3\gamma^2)}{1+3\gamma^2}\zeta^{-2}_\ast
   = \left\{ 
            \begin{array}{ll}
              2\zeta^{-2}_\ast & \textrm{\,\,\, for } \gamma=0, \\
               \zeta^{-2}_\ast & \textrm{\,\,\, for } \gamma=\frac{1}{3}.
            \end{array} 
      \right.
\label{eq:slope_condition}
\end{equation}  

\begin{figure}[t]
\includegraphics*[totalheight=0.85\columnwidth,width=\columnwidth]{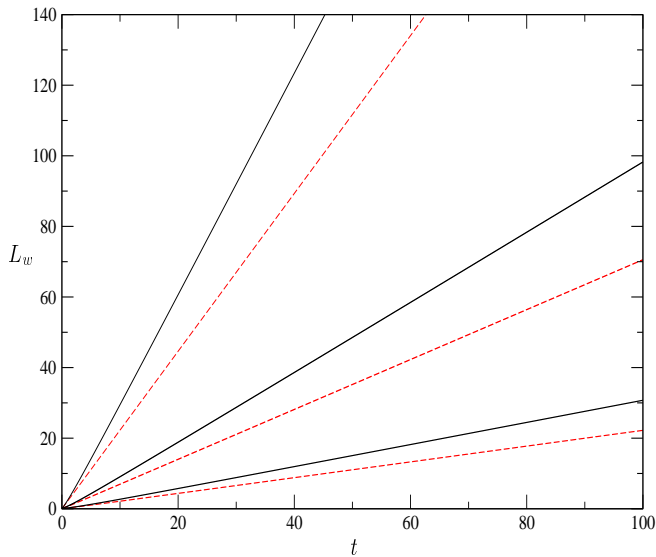}
\caption{Characteristic length of the winding mode network as a
function of cosmic time.
This graph shows the scaling behaviour of the decay
of winding modes.
The plotted curves represent solutions of the equations of motion 
with parameters $c$ and $\gamma$ chosen in the same manner as in 
Fig.~\ref{fig:phase_space}.
\label{fig:scaling}}
\end{figure}

\begin{table}
\caption{Slopes, $\zeta^{-1}_\ast$, of the characteristic length 
$L_w(t)$ plotted in Fig.~\ref{fig:scaling}.
We have also tabulated $\zeta^{-2}_\ast$ for an easy check of the
condition (\ref{eq:slope_condition}).
\label{tab:slopes}}
\begin{tabular}{*{5}{|c}|}
\cline{2-5}
\multicolumn{1}{c|}{}
& 
\multicolumn{2}{c|}{\tabequitem{$\gamma=0$}}
& 
\multicolumn{2}{c|}{\tabequitem{$\gamma=1/3$}}
\\
\cline{2-5}
\multicolumn{1}{c|}{}
&
\tabequitem{$\zeta^{-1}_\ast$}
&
\tabequitem{$\zeta^{-2}_\ast$}
&
\tabequitem{$\zeta^{-1}_\ast$}
&
\tabequitem{$\zeta^{-2}_\ast$}
\\
\hline
\, $c=0.1$\,
&
\, 0.2234500\,
&
\, 0.0499299\,
&
\, 0.3136377\,
&
\, 0.0983686\,
\\
\hline
\, $c=1.0$\,
&
\, 0.7069920\,
&
\, 0.4998377\,
&
\, 0.9945297\,
&
\, 0.9890893\,
\\
\hline
\, $c=10$\,
&
\, 2.2359990\,
&
\, 4.9996915\,
&
\, 3.1567680\,
&
\, 9.9651842\,
\\
\hline
\end{tabular}
\end{table}

{}From our numerical simulations (see Fig.~\ref{fig:scaling} and 
Table~\ref{tab:slopes}) we readily observed that $\zeta(t)$ relaxes very 
rapidly to a constant value and, then, $L_w$ scales with cosmic time $t$.
Moreover, apart from numerical errors, our solutions also satisfy the
above scaling relation with high accuracy.
To check if the system also scales with the cosmological horizon one 
has to recall from the previous subsection that at very early times
the size of the horizon can become extremely large but soon after the
winding modes have disappeared it behaves as, 
\begin{equation}
d_H
   \sim
l^{-1}
   \sim
\frac{1+3\gamma^2}{2\gamma}\, t.
\end{equation}
Consequently, this means that for values of $\gamma>0$ the characteristic 
length of the network configuration will inevitably evolve to a scale
comparable to the cosmological horizon.
On the contrary, if $\gamma=0$, the string network can never scale 
with $d_H$ despite it does with cosmic time.
This is closely related with the fact that, in this case, the universe 
reaches an indefinitely large period of loitering. 

We would also like to mention that we have check numerically that these 
scaling properties do not change significantly if we take the typical
size of the loops produced to scale with the characteristic length
of the winding mode network, $L_l \sim L_w$, instead of taking a size 
that scales with cosmic time.

\subsubsection{A gas of non-static branes}
The equation of state for the winding modes we have considered in 
our previous analysis, Eq.~ (\ref{eq:eq_state}), corresponds to 
a gas of non-relativistic branes. 
Let us briefly comment how we expect the cosmological dynamics will
be affected if the early evolution is dominated by a gas of 
non-static branes.
The generalised equation of state is characterised by a new 
parameter $\gamma_p$ given by (see for instance 
\cite{Boehm:2002bm,Vilenkin:1994}),
\begin{equation}
\gamma_p 
   = \left( \frac{p+1}{D}\langle v^2 \rangle
           -\frac{p}{D}
     \right),
\end{equation}
where $\langle v^2 \rangle$ is the average of the squared velocity
at each point of a brane.
For $\langle v^2 \rangle \rightarrow 1$ the gas of branes behaves
as a gas of relativistic particles whereas the non-relativistic or 
static limit corresponds to $\langle v^2 \rangle \rightarrow 0$.
In the case of winding modes with $p=1$ in $D=3$ spatial dimensions
we have a modified evolution equation for the Hubble parameter
$l$ depending on this new characteristic velocity of the brane gas, 
\begin{equation}
\dot l
   =  fl
     -\frac{1}{6}\,\mathrm{e}^\varphi 
      \left[ (1-2\langle v^2 \rangle ) E_w -3\gamma E_l
      \right]. 
\end{equation}
The first point to stress is that the very late-time cosmological 
evolution will not be qualitatively altered and the special lines
$l=-\gamma f$ will still continue to be solutions of the equations 
of motion and attractors of the rest of trajectories close to the 
critical point $(0,0)$ in the phase space spanned by $(f,l)$.
On the other hand, the rules for crossing these special lines
away from the origin are significantly modified and this has 
interesting dynamical consequences.
Eq~(\ref{eq:crossing_condition}) now reads,
\begin{equation}
\ddot\Phi
   = - \frac{\mathrm{e}^\varphi}{6}  
       \left( 1 + 3\gamma - 2\langle v^2 \rangle
       \right) E_w,
\end{equation}
and then, it is very easy to see, following the same argumentation 
of previous sections, that now crossings from values of $f$ and $l$ 
with $l+\gamma f<0$ to values with $l+\gamma f>0$ will be allowed if,
\begin{equation}
\langle v^2 \rangle
   > \frac{3\gamma+1}{2}.
\end{equation}
For $\gamma=0$ this means that the average velocity 
$\langle v^2 \rangle$ has to be greater than $1/2$.
However, if this condition is satisfied the Hubble parameter $l$ 
cannot have a turning point with $fl>0$, that is with a negative
value of $l$ (contracting phase).
Thus, contrary to what happens when the branes are static, a universe 
that was initially expanding will ever stay expanding.
But more interesting, if the universe started in a contracting phase 
it can now, for moderate values of the efficiency parameter, enter a 
stage of expansion before asymptotically reaching $(f,l)=(0,0)$.
This stage of late expansion when the branes of the gas are not 
static makes the case in which the loops produced behave like
ordinary matter to regain a phenomenological interest because  
this solves the obstruction discussed in previous sections to 
explain the dimensionality of our spacetime.

One might also ask whether it is still possible to ensure the
small string coupling approximation dynamically for solutions
with initial conditions satisfying the inequality $l<-f/3$ as
for the $\langle v^2 \rangle=0$ case since now crossings of the 
special lines are in principle allowed.
Fortunately, this criterion is still valid because the line $l+f/3=0$ 
cannot be cross from the region with $l+f/3<0$ (decreasing dilaton) 
to the region $l+f/3>0$ (growing dilaton) unless the brane gas has
an exotic characteristic velocity which exceeds the velocity of light,
$\langle v^2 \rangle > 1$.

With respect to the scaling properties of the winding mode decay,
we do not expect significative modifications relative to the picture
already outline for a gas of static branes.

\section{Conclusions\label{sec:Conclusions}}
In this work we have elaborated a detailed discussion of the
late-time behaviour of brane gas cosmologies which have served
to clarify and better understand some relevant issues of
their dynamics.

As a conclusion we have found that in order to obtain a 
phenomenologically interesting cosmological evolution the
loops produced without winding number have to behave as a 
gas with an equation of state characterised by a parameter 
satisfying $0<\gamma\leq 1/3$.
Otherwise, the string coupling becomes large and the low-energy 
effective description fails (for $\gamma > 1/3$), 
or the spatial dimensions do not grow and the dimensionality of 
our present universe cannot be explained (for $\gamma \leq 0$).
One would expect that the first difficulty could be alleviated 
simply by introducing higher-order quantum corrections.
As we have seen, one can avoid the second obstacle by considering 
the dynamics of a gas with non-static branes.
An alternative to obtain a late phase of expansion for 
$\gamma \leq 0$ is to include the effects of an axion or a moduli 
field.
The problem with this option is that the dilaton evolves in general
towards regions in phase space with a strong string coupling
\cite{Foffa:1999,Tsujikawa:2003pn}. 
In our analysis we have assumed for simplicity that the characteristic 
velocity of the brane gas $\langle v^2 \rangle$ is constant.
An interesting open question would be to find out whether the more 
realistic situation in which this velocity also varies with cosmic 
time could reveal new cosmological features. 

On the other hand, we have also confirmed that an early phase 
of loitering appears generically in brane gas scenarios if the 
efficiency parameter of the decay of winding modes takes moderate
values.
The existence of this phase in which the Hubble horizon becomes 
larger than the spatial extend of the universe offers a simple 
resolution to the {\sl brane problem} as suggested in 
\cite{Alexander:2000xv}.
In fact, the final fate of the universe in these type of cosmologies 
is always that of a loitering universe because the Hubble parameter 
goes to zero very slowly at late times.
Thus, even for those cases in which the early period of loitering 
does not occur, or is not long enough to provide a causal 
microphysical explanation for a complete disappearance of branes, 
the {\sl brane problem} can also be solved.

Finally, we have investigated whether the 
evolution of a network of winding modes driven by the 
dynamics of a dilaton field reaches a stage in which its 
characteristic length remains constant with respect to the size 
of the Hubble horizon $d_H$ as for ordinary cosmic strings in
a expanding universe.
We have shown that for values of $\gamma>0$ the characteristic 
length of the network will always scale with the cosmological 
horizon.
However, for the particular case in which $\gamma=0$, the network 
never scales with $d_H$ despite it does with cosmic time.
This is closely related with the fact that, in this case, the universe 
reaches an indefinitely large period of loitering.

\begin{acknowledgments}
The author thanks the support of the Alexander von Humboldt 
Foundation and the Universit\"at Heidelberg.
\end{acknowledgments}


\begin{thebibliography}{39}
\expandafter\ifx\csname natexlab\endcsname\relax\def\natexlab#1{#1}\fi
\expandafter\ifx\csname bibnamefont\endcsname\relax
  \def\bibnamefont#1{#1}\fi
\expandafter\ifx\csname bibfnamefont\endcsname\relax
  \def\bibfnamefont#1{#1}\fi
\expandafter\ifx\csname citenamefont\endcsname\relax
  \def\citenamefont#1{#1}\fi
\expandafter\ifx\csname url\endcsname\relax
  \def\url#1{\texttt{#1}}\fi
\expandafter\ifx\csname urlprefix\endcsname\relax\def\urlprefix{URL }\fi
\providecommand{\bibinfo}[2]{#2}
\providecommand{\eprint}[2][]{\url{#2}}

\bibitem[{\citenamefont{Brandenberger and Vafa}(1989)}]{Brandenberger:1989aj}
\bibinfo{author}{\bibfnamefont{R.~H.} \bibnamefont{Brandenberger}}
  \bibnamefont{and} \bibinfo{author}{\bibfnamefont{C.}~\bibnamefont{Vafa}},
  \bibinfo{journal}{Nucl. Phys.} \textbf{\bibinfo{volume}{B316}},
  \bibinfo{pages}{391} (\bibinfo{year}{1989}).

\bibitem[{\citenamefont{Tseytlin and Vafa}(1992)}]{Tseytlin:1992xk}
\bibinfo{author}{\bibfnamefont{A.~A.} \bibnamefont{Tseytlin}} \bibnamefont{and}
  \bibinfo{author}{\bibfnamefont{C.}~\bibnamefont{Vafa}},
  \bibinfo{journal}{Nucl. Phys.} \textbf{\bibinfo{volume}{B372}},
  \bibinfo{pages}{443} (\bibinfo{year}{1992}).

\bibitem[{\citenamefont{Tseytlin}(1992)}]{Tseytlin:1992ss}
\bibinfo{author}{\bibfnamefont{A.~A.} \bibnamefont{Tseytlin}},
  \bibinfo{journal}{Class. Quant. Grav.} \textbf{\bibinfo{volume}{9}},
  \bibinfo{pages}{979} (\bibinfo{year}{1992}).

\bibitem[{\citenamefont{Sakellariadou}(1996)}]{Sakellariadou:1996vk}
\bibinfo{author}{\bibfnamefont{M.}~\bibnamefont{Sakellariadou}},
  \bibinfo{journal}{Nucl. Phys.} \textbf{\bibinfo{volume}{B468}},
  \bibinfo{pages}{319} (\bibinfo{year}{1996}).

\bibitem[{\citenamefont{Cleaver and Rosenthal}(1995)}]{Cleaver:1995bw}
\bibinfo{author}{\bibfnamefont{G.~B.} \bibnamefont{Cleaver}} \bibnamefont{and}
  \bibinfo{author}{\bibfnamefont{P.~J.} \bibnamefont{Rosenthal}},
  \bibinfo{journal}{Nucl. Phys.} \textbf{\bibinfo{volume}{B457}},
  \bibinfo{pages}{621} (\bibinfo{year}{1995}).

\bibitem[{\citenamefont{Bassett et~al.}(2003)\citenamefont{Bassett, Borunda,
  Serone, and Tsujikawa}}]{Bassett:2003ck}
\bibinfo{author}{\bibfnamefont{B.~A.} \bibnamefont{Bassett}},
  \bibinfo{author}{\bibfnamefont{M.}~\bibnamefont{Borunda}},
  \bibinfo{author}{\bibfnamefont{M.}~\bibnamefont{Serone}}, \bibnamefont{and}
  \bibinfo{author}{\bibfnamefont{S.}~\bibnamefont{Tsujikawa}}
  (\bibinfo{year}{2003}), \eprint{hep-th/0301180}.

\bibitem[{\citenamefont{Alexander et~al.}(2000)\citenamefont{Alexander,
  Brandenberger, and Easson}}]{Alexander:2000xv}
\bibinfo{author}{\bibfnamefont{S.}~\bibnamefont{Alexander}},
  \bibinfo{author}{\bibfnamefont{R.~H.} \bibnamefont{Brandenberger}},
  \bibnamefont{and} \bibinfo{author}{\bibfnamefont{D.}~\bibnamefont{Easson}},
  \bibinfo{journal}{Phys. Rev.} \textbf{\bibinfo{volume}{D62}},
  \bibinfo{pages}{103509} (\bibinfo{year}{2000}).

\bibitem[{\citenamefont{Polchinski}(1995)}]{Polchinski:1995mt}
\bibinfo{author}{\bibfnamefont{J.}~\bibnamefont{Polchinski}},
  \bibinfo{journal}{Phys. Rev. Lett.} \textbf{\bibinfo{volume}{75}},
  \bibinfo{pages}{4724} (\bibinfo{year}{1995}).

\bibitem[{\citenamefont{Polchinski}(1996)}]{Polchinski:1996na}
\bibinfo{author}{\bibfnamefont{J.}~\bibnamefont{Polchinski}}
  (\bibinfo{year}{1996}), \eprint{hep-th/9611050}.

\bibitem[{\citenamefont{Polchinski}(1998)}]{Polchinski:1998}
\bibinfo{author}{\bibfnamefont{J.}~\bibnamefont{Polchinski}},
  \emph{\bibinfo{title}{String theory}} (\bibinfo{publisher}{Cambridge
  University Press}, \bibinfo{address}{Cambridge}, \bibinfo{year}{1998}).

\bibitem[{\citenamefont{Sen}(1996)}]{Sen:1996cf}
\bibinfo{author}{\bibfnamefont{A.}~\bibnamefont{Sen}}, \bibinfo{journal}{Mod.
  Phys. Lett.} \textbf{\bibinfo{volume}{A11}}, \bibinfo{pages}{827}
  (\bibinfo{year}{1996}).

\bibitem[{\citenamefont{Boehm and Brandenberger}(2002)}]{Boehm:2002bm}
\bibinfo{author}{\bibfnamefont{T.}~\bibnamefont{Boehm}} \bibnamefont{and}
  \bibinfo{author}{\bibfnamefont{R.}~\bibnamefont{Brandenberger}}
  (\bibinfo{year}{2002}), \eprint{hep-th/0208188}.

\bibitem[{\citenamefont{Easson}(2001)}]{Easson:2001fy}
\bibinfo{author}{\bibfnamefont{D.~A.} \bibnamefont{Easson}}
  (\bibinfo{year}{2001}), \eprint{hep-th/0110225}.

\bibitem[{\citenamefont{Easther
  et~al.}(2002{\natexlab{a}})\citenamefont{Easther, Greene, and
  Jackson}}]{Easther:2002mi}
\bibinfo{author}{\bibfnamefont{R.}~\bibnamefont{Easther}},
  \bibinfo{author}{\bibfnamefont{B.~R.} \bibnamefont{Greene}},
  \bibnamefont{and} \bibinfo{author}{\bibfnamefont{M.~G.}
  \bibnamefont{Jackson}}, \bibinfo{journal}{Phys. Rev.}
  \textbf{\bibinfo{volume}{D66}}, \bibinfo{pages}{023502}
  (\bibinfo{year}{2002}{\natexlab{a}}).

\bibitem[{\citenamefont{Watson and Brandenberger}(2003)}]{Watson:2002nx}
\bibinfo{author}{\bibfnamefont{S.}~\bibnamefont{Watson}} \bibnamefont{and}
  \bibinfo{author}{\bibfnamefont{R.~H.} \bibnamefont{Brandenberger}},
  \bibinfo{journal}{Phys. Rev.} \textbf{\bibinfo{volume}{D67}},
  \bibinfo{pages}{043510} (\bibinfo{year}{2003}).

\bibitem[{\citenamefont{Easther
  et~al.}(2002{\natexlab{b}})\citenamefont{Easther, Greene, Jackson, and
  Kabat}}]{Easther:2002qk}
\bibinfo{author}{\bibfnamefont{R.}~\bibnamefont{Easther}},
  \bibinfo{author}{\bibfnamefont{B.~R.} \bibnamefont{Greene}},
  \bibinfo{author}{\bibfnamefont{M.~G.} \bibnamefont{Jackson}},
  \bibnamefont{and} \bibinfo{author}{\bibfnamefont{D.}~\bibnamefont{Kabat}}
  (\bibinfo{year}{2002}{\natexlab{b}}), \eprint{hep-th/0211124}.

\bibitem[{\citenamefont{Alexander}(2002)}]{Alexander:2002gj}
\bibinfo{author}{\bibfnamefont{S.~H.~S.} \bibnamefont{Alexander}}
  (\bibinfo{year}{2002}), \eprint{hep-th/0212151}.

\bibitem[{\citenamefont{Zeldovich et~al.}(1974)\citenamefont{Zeldovich,
  Kobzarev, and Okun}}]{Zeldovich:1974uw}
\bibinfo{author}{\bibfnamefont{Y.~B.} \bibnamefont{Zeldovich}},
  \bibinfo{author}{\bibfnamefont{I.~Y.} \bibnamefont{Kobzarev}},
  \bibnamefont{and} \bibinfo{author}{\bibfnamefont{L.~B.} \bibnamefont{Okun}},
  \bibinfo{journal}{Zh. Eksp. Teor. Fiz.} \textbf{\bibinfo{volume}{67}},
  \bibinfo{pages}{3} (\bibinfo{year}{1974}), \bibinfo{note}{[Sov. Phys. JETP
  {\bf 40}, 1 (1974)]}.

\bibitem[{\citenamefont{Vilenkin and Shellard}(1994)}]{Vilenkin:1994}
\bibinfo{author}{\bibfnamefont{A.}~\bibnamefont{Vilenkin}} \bibnamefont{and}
  \bibinfo{author}{\bibfnamefont{E.~P.~S.} \bibnamefont{Shellard}},
  \emph{\bibinfo{title}{Cosmic string and other topological defects}}
  (\bibinfo{publisher}{Cambridge University Press},
  \bibinfo{address}{Cambridge}, \bibinfo{year}{1994}).

\bibitem[{\citenamefont{Brandenberger et~al.}(2002)\citenamefont{Brandenberger,
  Easson, and Kimberly}}]{Brandenberger:2001kj}
\bibinfo{author}{\bibfnamefont{R.}~\bibnamefont{Brandenberger}},
  \bibinfo{author}{\bibfnamefont{D.~A.} \bibnamefont{Easson}},
  \bibnamefont{and} \bibinfo{author}{\bibfnamefont{D.}~\bibnamefont{Kimberly}},
  \bibinfo{journal}{Nucl. Phys.} \textbf{\bibinfo{volume}{B623}},
  \bibinfo{pages}{421} (\bibinfo{year}{2002}).

\bibitem[{\citenamefont{Lema{\^\i}tre}(1931)}]{Lemaitre:1931}
\bibinfo{author}{\bibfnamefont{A.~G.} \bibnamefont{Lema{\^\i}tre}},
  \bibinfo{journal}{Mon. Not. R. astr. Soc.} \textbf{\bibinfo{volume}{91}},
  \bibinfo{pages}{483} (\bibinfo{year}{1931}).

\bibitem[{\citenamefont{Glanfield}(1966)}]{Glanfield:1966}
\bibinfo{author}{\bibfnamefont{J.~R.} \bibnamefont{Glanfield}},
  \bibinfo{journal}{Mon. Not. R. astr. Soc.} \textbf{\bibinfo{volume}{131}},
  \bibinfo{pages}{271} (\bibinfo{year}{1966}).

\bibitem[{\citenamefont{Felten and Isaacman}(1986)}]{Felten:1986}
\bibinfo{author}{\bibfnamefont{J.~E.} \bibnamefont{Felten}} \bibnamefont{and}
  \bibinfo{author}{\bibfnamefont{R.}~\bibnamefont{Isaacman}},
  \bibinfo{journal}{Rev. Mod. Phys.} \textbf{\bibinfo{volume}{58}},
  \bibinfo{pages}{689} (\bibinfo{year}{1986}).

\bibitem[{\citenamefont{Sahni et~al.}(1992)\citenamefont{Sahni, Feldman, and
  Stebbins}}]{Sahni:1991ks}
\bibinfo{author}{\bibfnamefont{V.}~\bibnamefont{Sahni}},
  \bibinfo{author}{\bibfnamefont{H.}~\bibnamefont{Feldman}}, \bibnamefont{and}
  \bibinfo{author}{\bibfnamefont{A.}~\bibnamefont{Stebbins}},
  \bibinfo{journal}{Astrophys. J.} \textbf{\bibinfo{volume}{385}},
  \bibinfo{pages}{1} (\bibinfo{year}{1992}).

\bibitem[{\citenamefont{Feldman and Evrard}(1993)}]{Feldman:1993ue}
\bibinfo{author}{\bibfnamefont{H.~A.} \bibnamefont{Feldman}} \bibnamefont{and}
  \bibinfo{author}{\bibfnamefont{A.~E.} \bibnamefont{Evrard}},
  \bibinfo{journal}{Int. J. Mod. Phys.} \textbf{\bibinfo{volume}{D2}},
  \bibinfo{pages}{113} (\bibinfo{year}{1993}), \eprint{astro-ph/9212002}.

\bibitem[{\citenamefont{Leigh}(1989)}]{Leigh:1989jq}
\bibinfo{author}{\bibfnamefont{R.~G.} \bibnamefont{Leigh}},
  \bibinfo{journal}{Mod. Phys. Lett.} \textbf{\bibinfo{volume}{A4}},
  \bibinfo{pages}{2767} (\bibinfo{year}{1989}).

\bibitem[{\citenamefont{Maggiore and Riotto}(1999)}]{Maggiore:1998cz}
\bibinfo{author}{\bibfnamefont{M.}~\bibnamefont{Maggiore}} \bibnamefont{and}
  \bibinfo{author}{\bibfnamefont{A.}~\bibnamefont{Riotto}},
  \bibinfo{journal}{Nucl. Phys.} \textbf{\bibinfo{volume}{B548}},
  \bibinfo{pages}{427} (\bibinfo{year}{1999}).

\bibitem[{\citenamefont{Veneziano}(1991)}]{Veneziano:1991ek}
\bibinfo{author}{\bibfnamefont{G.}~\bibnamefont{Veneziano}},
  \bibinfo{journal}{Phys. Lett.} \textbf{\bibinfo{volume}{B265}},
  \bibinfo{pages}{287} (\bibinfo{year}{1991}).

\bibitem[{\citenamefont{Gasperini and Veneziano}(2003)}]{Gasperini:2002bn}
\bibinfo{author}{\bibfnamefont{M.}~\bibnamefont{Gasperini}} \bibnamefont{and}
  \bibinfo{author}{\bibfnamefont{G.}~\bibnamefont{Veneziano}},
  \bibinfo{journal}{Phys. Rept.} \textbf{\bibinfo{volume}{373}},
  \bibinfo{pages}{1} (\bibinfo{year}{2003}).

\bibitem[{\citenamefont{Tsujikawa}(2003)}]{Tsujikawa:2003pn}
\bibinfo{author}{\bibfnamefont{S.}~\bibnamefont{Tsujikawa}}
  (\bibinfo{year}{2003}), \eprint{hep-th/0302181}.

\bibitem[{\citenamefont{Bennett}(1986{\natexlab{a}})}]{Bennett:1986qt}
\bibinfo{author}{\bibfnamefont{D.~P.} \bibnamefont{Bennett}},
  \bibinfo{journal}{Phys. Rev.} \textbf{\bibinfo{volume}{D33}},
  \bibinfo{pages}{872} (\bibinfo{year}{1986}{\natexlab{a}}).

\bibitem[{\citenamefont{Bennett}(1986{\natexlab{b}})}]{Bennett:1986zn}
\bibinfo{author}{\bibfnamefont{D.~P.} \bibnamefont{Bennett}},
  \bibinfo{journal}{Phys. Rev.} \textbf{\bibinfo{volume}{D34}},
  \bibinfo{pages}{3592} (\bibinfo{year}{1986}{\natexlab{b}}).

\bibitem[{\citenamefont{Brandenberger}(1994)}]{Brandenberger:1994by}
\bibinfo{author}{\bibfnamefont{R.~H.} \bibnamefont{Brandenberger}},
  \bibinfo{journal}{Int. J. Mod. Phys.} \textbf{\bibinfo{volume}{A9}},
  \bibinfo{pages}{2117} (\bibinfo{year}{1994}).

\bibitem[{\citenamefont{Campos}(2003)}]{Campos:2003}
\bibinfo{author}{\bibfnamefont{A.}~\bibnamefont{Campos}}, \bibinfo{journal}{in
  preparation}  (\bibinfo{year}{2003}).

\bibitem[{\citenamefont{Gerald and Wheatley}(1984)}]{Gerald:1984}
\bibinfo{author}{\bibfnamefont{C.~F.} \bibnamefont{Gerald}} \bibnamefont{and}
  \bibinfo{author}{\bibfnamefont{P.~O.} \bibnamefont{Wheatley}},
  \emph{\bibinfo{title}{Applied numerical analysis}}
  (\bibinfo{publisher}{Addison-Wesley Publishing Company},
  \bibinfo{address}{Reading}, \bibinfo{year}{1984}).

\bibitem[{\citenamefont{Bennett and Bouchet}(1988)}]{Bennett:1988vf}
\bibinfo{author}{\bibfnamefont{D.~P.} \bibnamefont{Bennett}} \bibnamefont{and}
  \bibinfo{author}{\bibfnamefont{F.~R.} \bibnamefont{Bouchet}},
  \bibinfo{journal}{Phys. Rev. Lett.} \textbf{\bibinfo{volume}{60}},
  \bibinfo{pages}{257} (\bibinfo{year}{1988}).

\bibitem[{\citenamefont{Bennett and Bouchet}(1989)}]{Bennett:1989ak}
\bibinfo{author}{\bibfnamefont{D.~P.} \bibnamefont{Bennett}} \bibnamefont{and}
  \bibinfo{author}{\bibfnamefont{F.~R.} \bibnamefont{Bouchet}},
  \bibinfo{journal}{Phys. Rev. Lett.} \textbf{\bibinfo{volume}{63}},
  \bibinfo{pages}{2776} (\bibinfo{year}{1989}).

\bibitem[{\citenamefont{Bennett and Bouchet}(1990)}]{Bennett:1990yp}
\bibinfo{author}{\bibfnamefont{D.~P.} \bibnamefont{Bennett}} \bibnamefont{and}
  \bibinfo{author}{\bibfnamefont{F.~R.} \bibnamefont{Bouchet}},
  \bibinfo{journal}{Phys. Rev.} \textbf{\bibinfo{volume}{D41}},
  \bibinfo{pages}{2408} (\bibinfo{year}{1990}).

\bibitem[{\citenamefont{Foffa et~al.}(1999)\citenamefont{Foffa, Maggiore, and
  Sturani}}]{Foffa:1999}
\bibinfo{author}{\bibfnamefont{S.}~\bibnamefont{Foffa}},
  \bibinfo{author}{\bibfnamefont{M.}~\bibnamefont{Maggiore}}, \bibnamefont{and}
  \bibinfo{author}{\bibfnamefont{R.}~\bibnamefont{Sturani}},
  \bibinfo{journal}{Nucl. Phys.} \textbf{\bibinfo{volume}{552}},
  \bibinfo{pages}{395} (\bibinfo{year}{1999}).

\end{thebibliography}
\end{document}